\documentclass{optica-article}

\journal{opticajournal} 

\articletype{Research Article}

\usepackage{lineno}

\begin{document}

\title{Ultrafast and compact photonic-electronic leaky integrate-and-fire circuits based upon resonant tunnelling diodes}

\author{Joshua Robertson,\authormark{1,*} Dylan Black,\authormark{1} Giovanni Donati,\authormark{1} Qusay Raghib Ali Al-Taai,\authormark{2} Ekaterina Malysheva,\authormark{3} Bruno Romeira,\authormark{4}  Jos\'{e} Figueiredo,\authormark{5} Victor Dolores Calzadilla,\authormark{3} Edward Wasige,\authormark{2}  and Antonio Hurtado\authormark{1}}

\address{\authormark{1}Institute of Photonics, SUPA Dept. of Physics, University of Strathclyde, Glasgow, UK\\
\authormark{2}High Frequency Electronics Group, University of Glasgow, Glasgow, UK\\
\authormark{3}Eindhoven Hendrik Casimir Institute, Eindhoven University of Technology, Eindhoven, Netherlands\\
\authormark{4}International Iberian Nanotechnology Laboratory, Braga, Portugal\\
\authormark{5}LIP and Departamento de Física da Faculdade de Ciências da Universidade de Lisboa, Lisboa, Portugal}

\email{\authormark{*}joshua.robertson@strath.ac.uk} 


\begin{abstract*} 
This work provides a first report of ultrafast and compact photonic-electronic neuromorphic temporal leaky integrate-and-fire neuronal circuits built with Resonant Tunnelling Diodes (RTDs). We demonstrate experimentally that multiple fast ($\sim$100-200\,ps) optical input pulses, arriving within a short (sub-ns long) temporal window, control the triggering of excitable responses in two different photonic-electronic RTD circuit architectures. These architectures are an electronic RTD coupled externally to a photodetector (referred to as a PD-RTD), and an integrated opto-electronic RTD device with inherent photodetection capability at infrared telecom wavelengths (referred to as an Optical RTD-PD). For both RTD systems, we reveal that the high-speed optically-triggered integrate-and-fire spiking operation can be precisely controlled by acting on the voltage bias applied to the RTD devices, or via the intensity of incoming optical pulses.  Finally, we demonstrate the application of the leaky integrate-and-fire behaviour to a pattern recognition task at high-speed, with the systems triggering fast ns-long electrical spikes in response to optical inputs of weighted 4-bit digital headers.

\end{abstract*}

\section{Introduction}

Neuromorphic computing, the emulation of biological neural functionality and neural circuits, has in recent years emerged as one of the key solutions for the next generation of computing architectures \cite{Frenkel2023}. Driven by the desire to improve substantially energy efficiency and achieve processing at the edge, neuromorphic systems enable the implementation of artificial neural networks, and in many cases Spiking Neural Networks (SNNs), using numerous electronic and photonic technologies \cite{Christensen2022}. Devices based on memristors \cite{Peng2024,Drouhin2022}, phase-change materials \cite{Rios2019,Feldmann2021}, modulators \cite{Xu2019,George2019} and semiconductors lasers \cite{Skalli2022,Zheng2023} (amongst others), are all seeing success by exploiting the nonlinear processes (all-optical, all-electrical or opto-electronic) in their approach, generating a plethora of neuromorphic contenders with a variety of benefits, timescales, and use-cases (e.g. image classification or real-time audio recognition) \cite{Srouji2023}. 

One type of promising neuromorphic device is Resonant Tunnelling Diodes (RTDs). RTDs are chip-scale semiconductor devices that are rapidly developing as candidates for high-speed and efficient artificial opto-electronic spiking neurons \cite{Ortega-Piwonka2021,Hejda2022,Hejda2023}. RTDs contain embedded double barrier quantum wells (DBQW) that give rise to highly non-linear I-V characteristics, producing regions of negative differential resistance (NDR) and different complex nonlinear dynamics, including crucially excitability and bursting \cite{Ortega-Piwonka2021,Hejda2022,Hejda2023}, enabling spike firing responses analogous biological neurons. RTDs have also demonstrated high bandwidth operation for applications such as THz oscillation generation and detection \cite{Izumi2017,Nishida2019}, thus offering promise for the realisation of ultrafast artificial spiking neuron operation. Further, opto-electronic RTDs, fabricated with embedded photodetecting layers, have shown the ability to deterministically trigger spiking regimes when subject to optical stimulation \cite{Pfenning2022,Al-Taai2023}. The versatility of RTDs, whether they are operated purely electronically or opto-electronically \cite{Zhang2024}, makes them suitable candidates for scalable, multimodal, non-linear spiking systems for present and future neuromorphic computing hardware.

\begin{figure}[htbp]
    \centering\includegraphics[width=11cm]{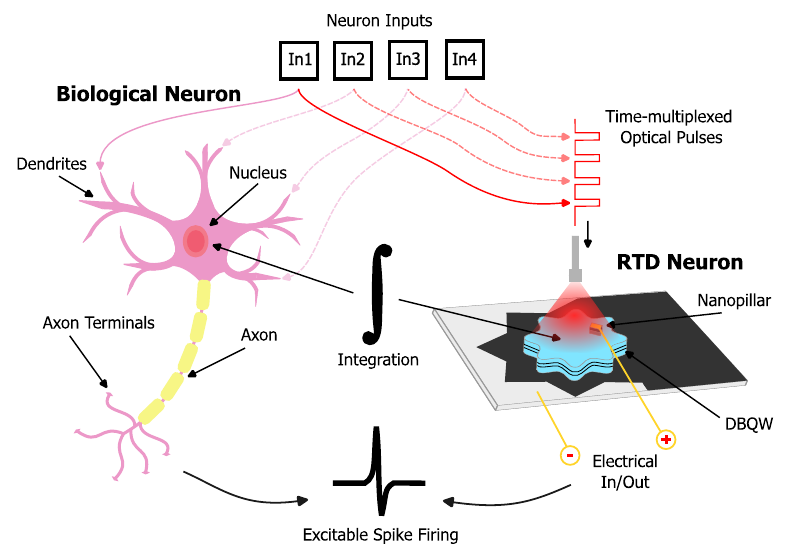}
    \caption{Schematic of a leaky integrate-and-fire neuron model and an optical RTD-PD leaky integrate-and-fire neuron. In the optical RTD-PD neuron multiple time-multiplexed optical inputs arrive on the neurons where temporal integration occurs within the device and an electrical spike is fired.}
    \label{fig:figure0}
\end{figure}

However, in SNN architectures several neuronal models can be implemented, each with different toolkits of functionality. One of these neuronal models, most commonly used for spike-based processing, is the leaky integrate-and-fire neuronal model \cite{Izhikevich2004}. These neurons are characterised by their ability to collect and combine input signals over time, with a counteracting 'leaky' decay constant, until the total contribution crosses a firing threshold and produces an excitable response. Leaky integrate-and-fire neurons, like those shown in \figurename{} \ref{fig:figure0}, therefore allow multiple inputs (temporal and spatial) from different dendritic pathways to contribute to the same spike activation, given the inputs occur within a temporal window shorter than the decay time of the system. Leaky integrate-and-fire neurons directly allow for the processing of input information, and for this reason several neuromorphic devices have seen investigations into this behaviour \cite{Drouhin2022,Zheng2023,Nahmias2013,Robertson2022}.

In this work, we demonstrate for the first time photonic-electronic RTD neuronal circuits exhibiting a temporal leaky integrate-and-fire behaviour when subject to time-multiplexed optical pulses (see \figurename{} \ref{fig:figure0}). Two RTD architectures are investigated, which include an electronic RTD externally-coupled to a fast photodetector, and an opto-electronic RTD with an embedded infrared light detection semiconductor layer. In this work, both RTD circuits are applied to a practical 4-bit pattern recognition task to showcase their potential as scalable opto-electronic devices for neuromorphic systems.   

\section{Methods}

\begin{figure}[htbp]
    \centering\includegraphics[width=11cm]{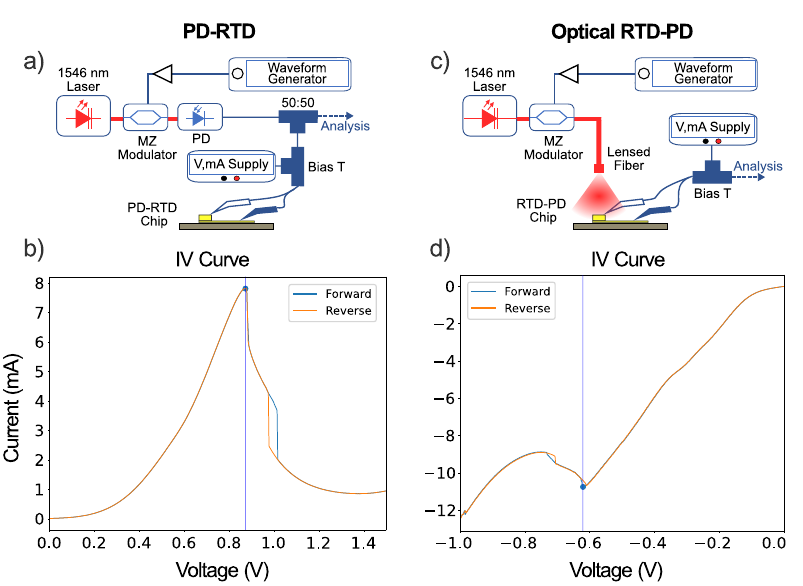}
    \caption{Experimental setup diagrams and characteristic I-V curves of both the PD-RTD (a-b) and the optical RTD-PD (c-d) systems investigated in this work. The 'peak' voltages at the onset of the NDR (blue line) for the PD-RTD and optical RTD-PD are 0.879\,V and -0.623\,V, respectively.}
    \label{fig:figure1}
\end{figure}

Two types of neuromorphic spiking RTD structures were investigated in this work. The first structure is a 3\,$\mu$m radius mesa, fully electronic, RTD featuring an AlAs/InGaAs/AlAs double barrier quantum well (DBQW) \cite{Hejda2023}. It is fed both a DC and an RF electrical input (via RF probes and a bias tee) that arrive from a power supply and a fast amplified photodetector (PD), respectively. In this configuration (referred to here as a PD-RTD), optical inputs incident on the PD are converted to electrical signals for injection into the RTD device. The optical signals are generated by modulating the output of a 1546\,nm tuneable laser source with a Mach-Zehnder (MZ) intensity modulator and an Arbitrary Waveform Generator (AWG). A 50:50 splitter is used to analyse the RF output of the PD-RTD following the injection of input signals. A schematic of the PD-RTD setup is shown in \figurename{} \ref{fig:figure1}\,(a). The second structure studied in this work is an opto-electronic RTD with a similar DBQW stack, but with the inclusion of an embedded spacer layer (that enables light absorption) and top layer nanopillar structure (for improved carrier confinement) \cite{Al-Taai2023}. In this configuration (referred to here as an Optical RTD-PD), the modulated optical inputs from a 1546\,nm tuneable laser are directly introduced to the RTD device via a lensed fiber. The lensed fibre delivered an average power of 0.885\,mW to the light absorbing spacer layer of the RTD. The RTD device was driven with a DC bias from a power supply ($V_{bias}$), and again the RF output of the RTD was monitored for analysis. A schematic of the optical RTD-PD setup is shown in \figurename{} \ref{fig:figure1}\,(c). In both configurations a 12\,GSa/s (5\,GHz) arbitrary waveform generator was used to encode fast optical pulses in the system, and a fast 40\,GSa/s (16\,GHz) oscilloscope was used to analyse and capture the corresponding electrical responses. Full details of the structures, dimensions and epilayer structures of the RTD devices used in this work can be found in \cite{Hejda2023,Al-Taai2023}. 

The current-voltage (I-V) characteristics of both the PD-RTD and optical RTD-PD devices are shown in \figurename{} \ref{fig:figure1}\,(b) and \figurename{} \ref{fig:figure1}\,(d), respectively. These plots reveal the regions of negative differential resistance (NDR) that give rise to non-linear excitability (spiking) in the RTD devices \cite{Ortega-Piwonka2021}. To make this measurement, the bias voltage of the devices was swept to identify the 'peak' value at the onset of the NDR (blue line). The 'peak' of the PD-RTD was measured to be 0.879\,V and the 'peak' of the optical RTD-PD was measured to be -0.623\,V (in a reverse biased configuration). In this work, both devices were operated outside the NDR in the region near their 'peak' to allow incoming optical perturbations to deterministically trigger spiking responses: the PD-RTD (Optical RTD-PD) was typically supplied a DC bias of 0.863\,V (-0.611\,V) prior to light illumination.

\section{Integrate-and fire Operation with Temporal Optical Inputs}
\begin{figure}[htbp]
    \centering\includegraphics[width=11cm]{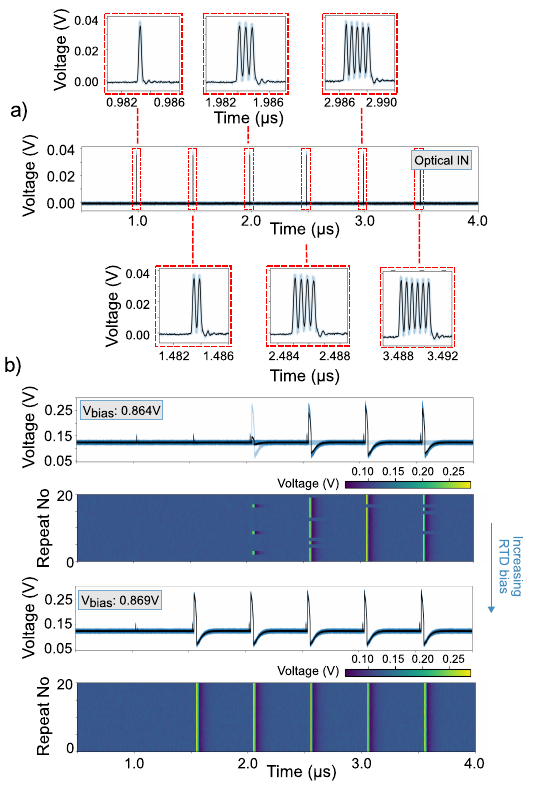}
    \caption{PD-RTD subject to bursts of optical input pulses. The optical input bursts (a) feature an increasing number of positive pulses with input amplitudes of 35\,mV that are 200\,ps-long with 200\,ps temporal spacing. Time-traces and consistency plots for the PD-RTD's output (b) are shown for two $V_{bias}$ conditions, 0.864\,V \& 0.869\,V.}
    \label{fig:figure2}
\end{figure}

In this work, we test each of the RTD systems with bursts of (sub-threshold) temporally separated inputs, occurring within a short temporal window, to investigate the existence of the leaky integrate-and-fire mechanism in the devices. 
The test for the leaky integrate-and-fire behaviour in the PD-RTD is shown in \figurename{} \ref{fig:figure2}. Six bursts of 35\,mV optical inputs are encoded and injected into the PD-RTD system, each below the threshold for spiking firing, with an increasing number of fast optical pulses (1 to 6 pulses). The bursts of optical inputs were separated by 500\,ns, and each contained 200\,ps-long optical pulses spaced by 200\,ps (as shown in \figurename{} \ref{fig:figure2}\,(a)). The time-traces shown in \figurename{} \ref{fig:figure2}\, (and throughout this work) plot in blue the overlap of all measured repetitions, with a typical single trace plotted in black. The optical inputs injected into the PD-RTD are positive pulses. Positive pulses shift the operating point of the PD-RTD towards the NDR region, allowing it to produce excitable responses when pushed beyond the 'peak' \cite{Hejda2023}. The electrical PD-RTD system responses are shown for several bias voltages ($V_{bias}$) in \figurename{} \ref{fig:figure2}\,(b). At a $V_{bias}$ of 0.864\,V, the input bursts of 1-3 optical pulses failed to consistently trigger excitable spiking responses in the PD-RTD. However, the inputs with more temporally-grouped pulses (4-6 pulses), successfully triggered excitable spiking responses with high consistency. The PD-RTD spiking responses (\figurename{} \ref{fig:figure2}\,(b)) occur in the form of positive fast-slow-fast-slow transitions, and in this device have been shown to activate with a refractory period of 90 ns \cite{Hejda2023}. The spiking responses for the $V_{bias}$ of 0.864\,V reveal not only the existence of an activation threshold in the PD-RTD system \cite{Hejda2023}, but that multiple sub-threshold inputs, despite arriving at different times, are integrated by the RTD towards a single spike firing event. Similarly, at the higher $V_{bias}$ of 0.869\,V (an operating point closer to the 'peak' value in the device's I-V curve), all bursts of temporally-grouped optical pulses triggered spiking responses, with only the single optical input pulse failing to elicit a response. This demonstrates that the $V_{bias}$ voltage can be precisely controlled at will to influence the threshold of the integrate-and-fire response. Here, through this test, we readily observe the capability of the system to integrate multiple temporal inputs, similar to a leaky integrate-and-fire neuronal model. Bursts with more input pulses demonstrated more reliable spiking further from the 'peak', as their total integrated input allowed them to overcome the larger threshold required for spike firing. We note that when operating with inputs of increased pulse spacing, the PD-RTD's integrate-and-fire functionality quickly broke down. This feature reveals the PD-RTD system operates with a 'leaky' temporal decay constant, meaning that optical inputs must occur within a short temporal window to integrate effectively. We believe the timescale of the integration window is correlated with that of fast rising edge transition of the PD-RTD spike, measured to be approximately 500\,ps. The integrate-and-fire behaviour of the PD-RTD neuron therefore enables operation and functionality with multiple GHz-rate, temporally-separated or time-multiplexed, optical inputs. 

\begin{figure}[htbp]
    \centering\includegraphics[width=11cm]{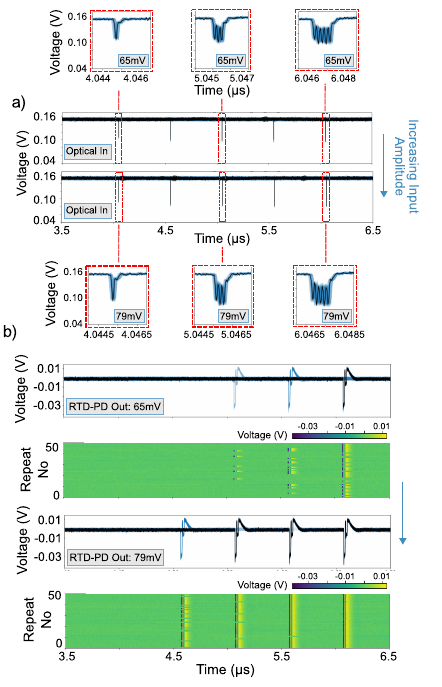}
    \caption{Optical RTD-PD subject to bursts of optical input pulses. The optical input bursts (a) feature an increasing number of negative optical input pulses that are 100\,ps-long with 100\,ps temporal spacing. Two sets of inputs are tested, one with input drop amplitudes of 65\,mV and one with drop amplitudes of 79\,mV. Time-traces and consistency plots for the optical RTD-PD's output (b) are shown for each amplitude case.}
    \label{fig:figure3}
\end{figure}

Similar optical input signals were used to test fast temporal integration responses in the optical RTD-PD system, as shown in \figurename{} \ref{fig:figure3}. Here, five bursts of (sub-threshold) optical input pulses were encoded as negative drops in the optical signal directly injected into the device, enabling spike triggering at the 'peak' with the optically-sensitive reverse-biased RTD-PD system \cite{Al-Taai2023}. The input optical pulses were generated at a higher speed, with a reduced pulse width of 100\,ps, and a narrower temporal spacing between consecutive pulses of 100\,ps. Alternatively, in this second demonstration the $V_{bias}$ voltage was kept constant, with now the modulation amplitude of the optical input signal being varied. The increased input signal amplitude placed across the MZ modulator produced different drop amplitudes of 65 and 79\,mV, when measured using a photodetector (see \figurename{} \ref{fig:figure3}\,(a)). The output time-traces and consistency plots of the optical RTD-PD, following direct light injection via the lensed-fibre, are shown in \figurename{} \ref{fig:figure3}\,(b). For the case of input drop amplitude of 65\,mV, we clearly see the existence of a spike activation threshold. Further, we achieve an increase in the consistency of spiking responses for the bursts with higher numbers of temporally-grouped input pulses. We note here that due to the reverse-biasing of the device, and the alternative spike-firing mechanism of the optical RTD-PD (the shift of the characteristic I-V curve when directly exposed to incident light, see \cite{Al-Taai2023}), the spiking responses emerge flipped, and demonstrate a larger refractory period of 120\,ns. When the input amplitudes are increased to 79\,mV, we achieved spike activation for all bursts of temporally-grouped pulses, with only the single inputs failing to trigger responses. Again, through these results we observe that controlling the amplitude of the optical inputs controls the activation threshold of the system, as now more spiking responses are generated despite the same $V_{bias}$. Further, we observe larger groups of identical sub-threshold temporal optical pulses producing spiking responses before smaller groups, a direct demonstration of the temporal integration towards a single firing event in the optical RTD-PD device. We note that similar to the PD-RTD neuron, increasing the temporal spacing of inputs beyond 500\,ps lead to the break down of the temporal integration. This again highlights the presence of a 'leaky' temporal decay in the optical RTD-PD neuron. Again, we observed that the timescale of the integration window correlated with fast falling edge transition of the spike, measured to be approximately 200\,ps. Both the integration window and the rise/fall time of the optical RTD-PD spike are narrower than the PD-RTD, enabling operation with higher-rate optical inputs. Specifically, we used 100\,ps-long optical pulses with 100\,ps spacing (due to experimental limitations) but operation with even faster optical input signals should be possible. We anticipate that the difference between the two RTD neurons is caused in part by the use of an external photodetector in the PD-RTD system. The external components and cabling may cause changes in the overall circuit equivalent inductance, altering the frequency of the RTD response, an effect also responsible for the MHz spiking rates observed in the system \cite{Hejda2023,Al-Taai2023}. Further, different device geometries and epilayer structures can lead to different rise/fall transition times during spiking. Given the similarity between the integration and transition timescales, these parameter may play a role in the rate at which integrate-and-fire operation can be achieved. Nevertheless, the integration of temporal-separated or time-multiplexed optical inputs is possible within optical RTD-PDs given inputs occur at fast GHz-rates, offering exciting prospects for fast RTD-based photonic-electronic neuromorphic processing platforms in the future.

\section{4-bit Pattern Recognition via Temporal Integration}

\begin{figure}[htbp]
    
    \centering\includegraphics[width=11cm]{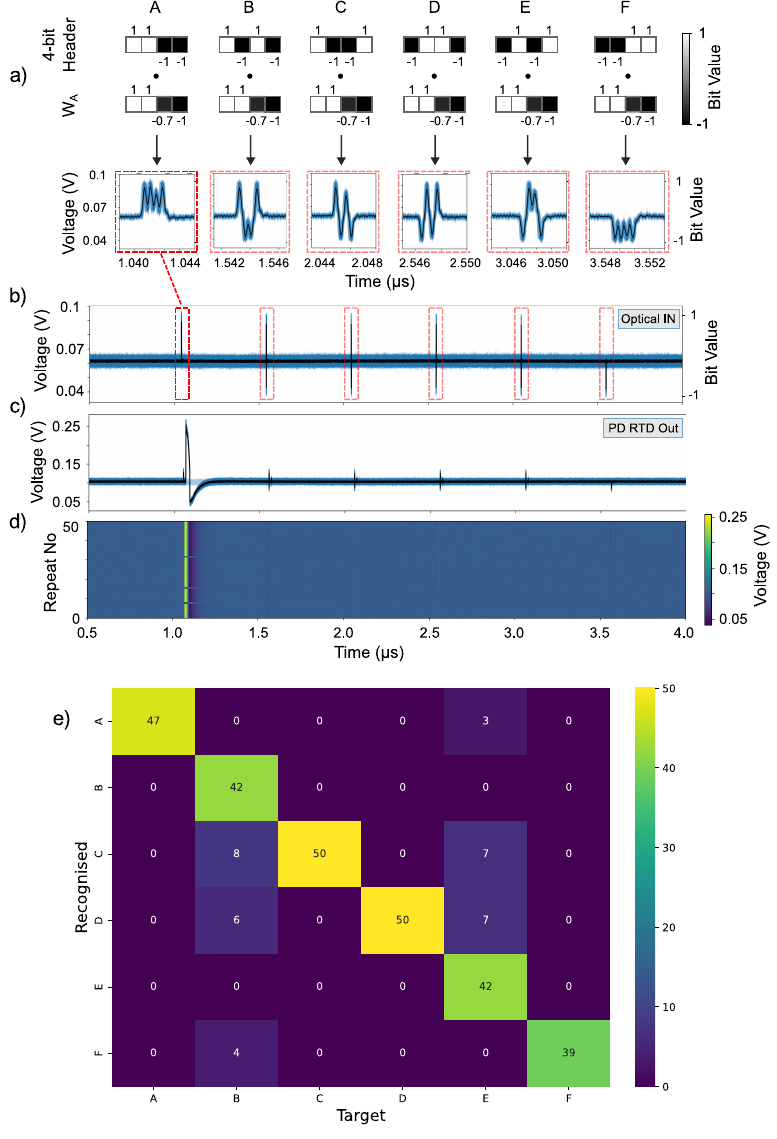}
    \caption{Experimental demonstration of the PD-RTD performing a 4-bit pattern recognition task with a leaky integrate-and-fire behaviour. Each 4-bit pattern, the applied weight matrix and the pre-processed weighted inputs are shown in (a \& b). The corresponding PD-RTD output time-trace (c) and consistency map (d) show successful activation for the target pattern (Pattern A - 1100). A confusion matrix shows performance for the recognition of all pattern across 50 repetitions, achieving a global accuracy of 88.5\%. Applied weight matrices: $W_{A}$(1,1,-0.7,-1), $W_{B}$(1,-1,1,-1), $W_{C}$(1,-1,-1,1), $W_{D}$(-1,1,1,-1), $W_{E}$(-1,0.9,-0.9,1), $W_{F}$(-1,-1,0.7,1).}
    \label{fig:figure4}
\end{figure}

\begin{figure}[htbp]
    
    \centering\includegraphics[width=11cm]{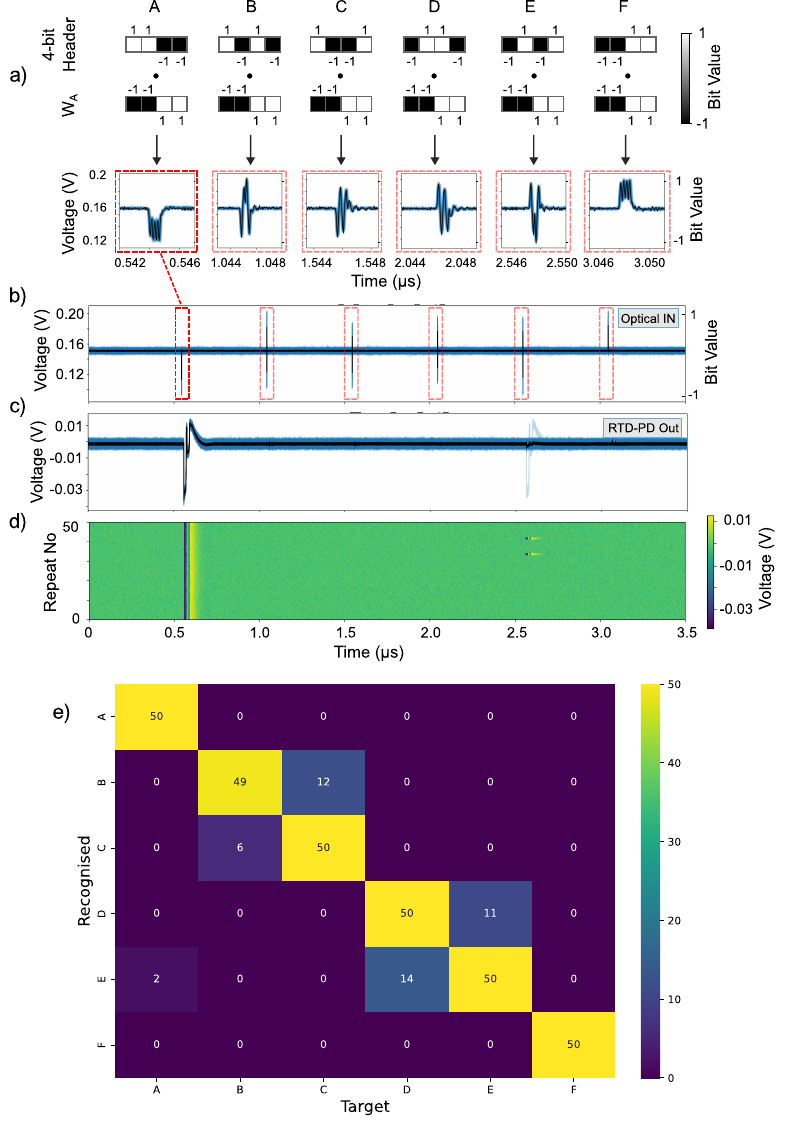}
    \caption{Experimental demonstration of the optical RTD-PD performing a 4-bit pattern recognition with a leaky integrate-and-fire behaviour. Each 4-bit pattern, the applied weight matrix and the pre-processed weighted inputs are shown in (a \& b). The corresponding optical RTD-PD output time-trace (c) and consistency map (d) show successful activation for the target pattern (Pattern A - 1100). A confusion matrix shows performance for the recognition of all pattern across 50 repetitions, achieving a global accuracy of 86.9\%. Applied weight matrices: $W_{A}$(1,1,-1,-1), $W_{B}$(1,-1,1,-1), $W_{C}$(1,-1,-1,1), $W_{D}$(-1,1,1,-1), $W_{E}$(-1,1,-1,1), $W_{F}$(-1,-1,1,1).}
    \label{fig:figure5}
\end{figure}

To demonstrate the processing capability of the investigated leaky integrate-and-fire neuronal behaviour in the PD-RTD and optical RTD-PD, we tested both systems on a 4-bit header recognition task at high-speed. Six different 4-bit headers (A - 1100, B - 1010, C - 1001, D - 0110, E - 0101 \& F - 0011) were first pre-processed offline, where each bit was assigned a $+/-$\,1 value and weighted by a weight matrix $W_{A-F}$. The weight matrix was programmed to enable recognition of a single 4-bit header from a line-up of all six consecutively-injected patterns. Following offline weighting, each bit of the weighted 4-bit headers was time-multiplexed into a high-speed burst of four temporally-grouped optical pulses. For the case of the PD-RTD (optical RTD-PD), the pulse amplitudes were programmed with the weighted bit values with pulse durations of 200\,ps (100\,ps) per pulse, and 200\,ps (100\,ps) spacing. The six 4-bit headers, the weight matrix, and the resulting optical inputs for the case of Pattern A (1100) recognition, are shown in \figurename{} \ref{fig:figure4}\,(a-b) and \figurename{} \ref{fig:figure5}\,(a-b). These plots correspond to the case of the PD-RTD and optical RTD-PD system, respectively. The weight matrix values used in \figurename{} \ref{fig:figure5} are flipped to correctly orientate the optical input pulses. From the time-traces and consistency maps measured at the output of the two RTD systems, shown respectively in \figurename{} \ref{fig:figure4}\,(c-d) and \figurename{} \ref{fig:figure5}\,(c-d), we see that both leaky integrate-and-fire systems produced spikes in response to the target, Pattern A - 1100, only. Each system responds for this pattern because the temporal integration of Pattern A's pulses generates the largest combined optical input. The RTD devices are therefore performing the function of a leaky integrate-and-fire neuron, taking incident weighted inputs, integrating them temporally, and producing the appropriate spiking response. The recognition of all targets was performed by re-configuring $W_{B-F}$, with the overall recognition performance plotted in the confusion matrices of \figurename{} \ref{fig:figure4}\,(e) and \figurename{} \ref{fig:figure5}\,(e) (see captions for $W_{B-F}$ values). Good overall performance was achieved in this task, with the PD-RTD yielding a global accuracy of 88.5\% and the optical RTD-PD producing a global accuracy of 86.9\%, despite no specific learning rules having been implemented in the selection of weights for this initial proof-of-concept demonstration. 

\section{Conclusion}
In conclusion, we demonstrate for the first time that photonic-electronic RTD neurons (including an electrical PD-RTD and an optical RTD-PD) have the capability to demonstrate leaky integrate-and-fire operation, similar to biological neuronal models, but at sub-nanosecond speeds. For the RTD systems investigated in this work, fast bursts of sub-threshold optical input pulses, arriving within short sub-ns temporal windows (much shorter than the refractory time of the RTD circuits), were able to trigger excitable spiking responses by integrating the contribution of each optical input over time. Further, we highlighted that the threshold for this leaky integrate-and-fire excitation can be fully and precisely controlled using either the RTD bias voltage or the amplitude of the optical inputs. Finally, leveraging the demonstrated leaky integrate-and-fire behaviour, we showed that both systems, the PD-RTD and the optical RTD-PD, yielded good performance in a 4-bit header recognition task with sub-ns bits. These demonstrations of neuromorphic behaviours open up exciting new avenues for RTD-based implementation in photonic-electronic spike-based processing platforms on chip, with operation in key infrared telecom wavelength windows for seamless integration with optical networks and data centre technologies, towards scalable brain-like processing hardware.  


\medskip
\begin{backmatter}
\bmsection{Funding}
The authors acknowledge support by the European Commission (Grant 828841-ChipAI-H2020-FETOPEN-2018-2020 and EIC Pathfinder ‘SpikePRO’ programme), and by the UK Research and Innovation (UKRI) Turing AI Acceleration Fellowships Programme (EP/V025198/1).

\bmsection{Acknowledgment}
The authors would like to acknowledge IQE plc. for providing the semiconductor wafers used to fabricate the systems of this work. 

\bmsection{Disclosures}
The authors declare no conflicts of interest.

\bmsection{Data Availability Statement}
For the purpose of open access, the author(s) has applied a Creative Commons Attribution (CC BY) licence to any Author Accepted Manuscript version arising from this submission. All data underpinning this publication are openly available from the University of Strathclyde KnowledgeBase at https://doi.org/10.15129/e2b593f1-fc26-4097-858d-62a1bef0a519

\end{backmatter}
\bibliography{Manuscript}

\end{document}